\documentclass{iau}

\usepackage{amsmath}
\usepackage{graphicx}
\usepackage{multirow}
\usepackage{natbib}
\usepackage{caption}

\begin{document}

\lefttitle{Jayanand Maurya}
\righttitle{Origin of the eMSTO in open cluster NGC 2355}

\jnlPage{1}{7}
\jnlDoiYr{2021}
\doival{10.1017/xxxxx}

\aopheadtitle{Proceedings IAU Symposium}
\editors{C. Sterken,  J. Hearnshaw \&  D. Valls-Gabaud, eds.}

\title{Unraveling eMSTO in NGC 2355: Stellar Rotation and Binaries}

\author{Jayanand Maurya, Yu Zhang, and Hubiao Niu}
\affiliation{Xinjiang Astronomical Observatory, Chinese Academy of Sciences, No. 150, Science 1 Street, Urumqi, Xinjiang 830011, People’s Republic of China}

\begin{abstract}
The extended Main Sequence Turn-off (eMSTO) in the open cluster NGC 2355 is investigated using precise astrometry and photometry from Gaia DR3 and spectroscopic data from the Gaia-ESO Survey. We find a clear positive correlation between the rotational velocity ($v \sin i$) and color of eMSTO stars, supporting the role of stellar rotation and gravity darkening in causing the observed color spread. Contrary to predictions from binary tidal-locking scenarios, spatial distribution analysis reveals that fast-rotating stars are preferentially concentrated in the cluster's central regions. This finding provides the observational evidence favoring star-disk interactions during the pre-main-sequence phase, rather than tidal locking in binaries, as the primary mechanism responsible for the origin of rotational spread in eMSTO stars.
\end{abstract}


\maketitle

\section{Introduction}

The unusual spread in the upper main sequence (MS) near turn-off in the color-magnitude diagram (CMD) of a open cluster is known as extended Main Sequence Turn-off, i.e. eMSTO. The eMSTO in star clusters of Magellanic Clouds galaxy are known to be commonly present but the discovery of the eMSTO in the Milky Way open clusters has been discovered recently by \cite{2018ApJ...863L..33M}. \cite{2009MNRAS.398L..11B} suggested that the spread in rotation rates of the stars can manifest into the spread in color of the MS stars near turn-off due to gravity darkening effect. The fast rotation causes reduced gravity near stellar equator due to centrifugal forces which results into smaller effective temperature and luminosity for the star \citep{1924MNRAS..84..665V}. The combination of viewing angles and gravity darkening can produce the observed color spread in the eMSTO stars of a open star cluster \citep{2024MNRAS.532.1212M}.    

Positive correlations have been found between observed color of the eMSTO stars and their projected rotational velocities, \textit{v}sin\textit{i} \citep{2020MNRAS.492.2177K,2025ApJ...989..123M}. Several mechanisms have been suggested for the production of the spread in rotation rates of the eMSTO stars with similar masses in an open cluster. \cite{2015MNRAS.453.2637D} proposed that tidal interactions in the binaries with different mass-ratio and orbital separation slow-down their rotation rates causing an spread in rotation rates. Alternatively, \cite{2020MNRAS.495.1978B} attribute rotational spreads to star-disk interactions during pre-main-sequence phase of stellar evolution. Stars losing their circumstellar disks early in the pre-main sequence lifetime spin up during contraction, while those retaining disks longer rotate slower. The first observational evidence supporting the star-disk interaction mechanism for the origin of the spread in rotation rates was found by \cite{2024MNRAS.532.1212M}. The degeneracy between star-disk interaction and tidal locking in binaries as the primary mechanism responsible for the rotational spread among eMSTO stars remains an open question, necessitating further study \citep{2025ApJ...989..123M}.

In this article, we present a brief discussion on the various aspects of the origin of the eMSTO in open cluster NGC 2355. A detailed investigation of the eMSTO phenomenon in NGC 2355 can be found in our previous article, \citet{2024MNRAS.532.1212M}.  

\section{Data and Membership}

We used precise photometric and astrometric data from \textit{Gaia} DR3 to identify member stars and estimate physical parameters of NGC 2355. \textit{Gaia} DR3 provides high-precision astrometry ($\sim$0.03 mas yr$^{-1}$ in proper motion, $\sim$0.03 mas in parallax) and photometry (uncertainties $\sim$0.001 mag in G-band) for stars brighter than 17 mag \citep{2023A&A...674A...1G}, enabling effective field star decontamination essential for eMSTO detection. For spectroscopic analysis, we utilized archived medium and high-resolution spectra from the Gaia–ESO Survey (ESO programme 197.B-1074), obtained with the GIRAFFE (R$\sim$19,200; 644-680~nm) and UVES (R$\sim$51,000; 582-683~nm) spectrographs. These spectra were used to determine the projected rotational velocities ($v \sin i$) of eMSTO stars.

We determined stellar membership by calculating probabilities through a statistical method based on proper motions of stars as described in \cite{2020MNRAS.494.4713M}. Stars were selected as cluster members if they had a membership probability exceeding 60\% and a parallax value within 3 standard deviations of the mean cluster parallax. This way, we found a total of 411 member stars in NGC 2355 \citep{2024MNRAS.532.1212M}. 

\section{Rotational properties of the eMSTO stars}
The projected rotational velocity, $v \sin i$, of the eMSTO stars were determined using the \textit{iSpec} software package \citep{2014A&A...569A.111B,2019MNRAS.486.2075B}. Our analysis employed the SPECTRUM radiative transfer code \citep{1994AJ....107..742G} together with MARCS model atmospheres \citep{2008A&A...486..951G}, both available within iSpec, to generate synthetic spectra. These models utilized solar abundances from \cite{2005MSAIS...8...14K}. The $v \sin i$ were determined by fitting synthetic spectra to the observed data. This was achieved by employing a global $\chi^2$ minimization technique to find the optimal match between the model and observed spectral features (see \cite{2024MNRAS.532.1212M} for detailed approach).

\begin{figure}
    \includegraphics[scale=.5]{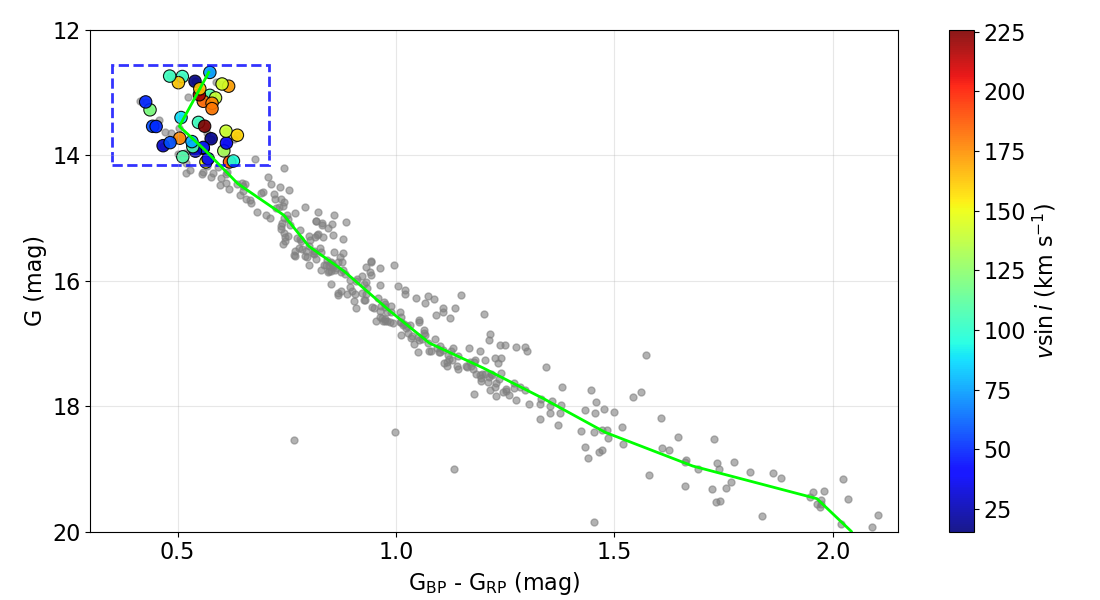}
    \caption{Color-magnitude diagram for NGC 2355. Gray points represent cluster members. eMSTO stars with measured $v \sin i$) are overplotted and color-coded by their $v \sin i$ values. The lime-green line shows the fiducial MS. The blue dashed rectangle highlights the eMSTO region in the upper MS.}
    \label{cmd_colorbar}
  \end{figure}

\begin{figure}
    \includegraphics[scale=.45]{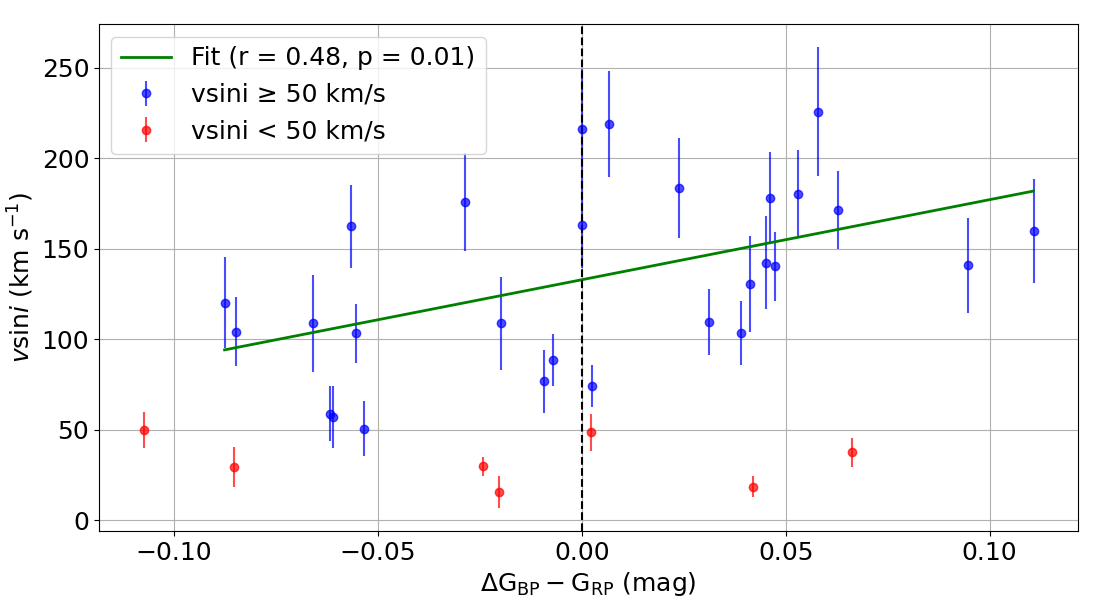}
    \caption{The correlation between pseudo color $\Delta$G$_{\rm BP}-G_{\rm RP}$ and $v \sin i$ for eMSTO stars. The red points ($v \sin i < 50$ km s$^{-1}$) are slow-rotating outliers.}
    \label{corr_vsini_color}
  \end{figure}

We found that fast-rotating stars are predominantly located on the red side, while slow-rotating stars are found on the blue side of the eMSTO region as visible in Figure~\ref{cmd_colorbar}. We derived fiducila MS line from median values of the bins with width 0.5 mag in G band. The mean rotational velocity for blue-eMSTO stars is 89.1 $\pm$ 4.5 km s$^{-1}$, compared to 138.1 $\pm$ 5.3 km s$^{-1}$ for the red-eMSTO stars. A clear positive correlation is observed between rotation rates and color of the eMSTO stars having a Pearson correlation coefficient of 0.48 with p-value = 0.01 as shown in Figure~\ref{corr_vsini_color}. This positive correlation indicates that stellar rotation due to gravity darkening might cause the observed color spread in the eMSTO population of NGC 2355. The outliers, marked in red, were excluded from the linear regression in Figure~\ref{corr_vsini_color}. These slowly rotating eMSTO stars are likely undetected binaries that could not be identified with the currently available single-epoch spectra.

\section{Physical mechanism causing the spread in rotation rates}
\begin{figure}
    \includegraphics[scale=.33]{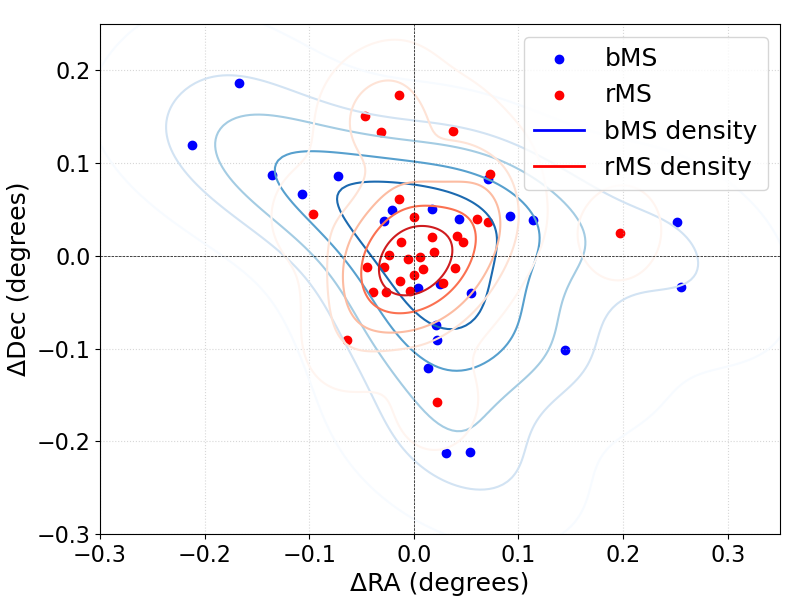}
    \includegraphics[scale=.33]{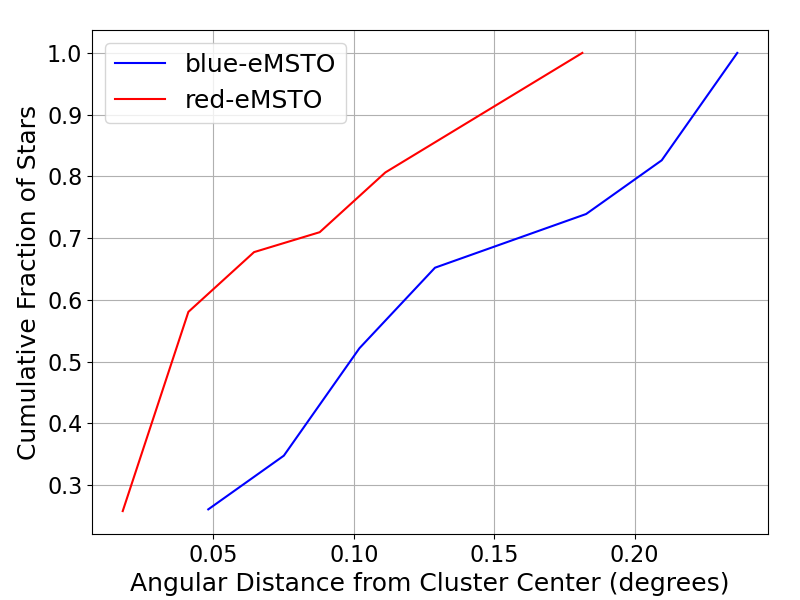}
    \caption{Left panel: Spatial distribution of blue-eMSTO and red-eMSTO main-sequence stars relative to the cluster center. The density contours illustrate the concentration of each stellar population. Right panel: Cumulative radial distributions of the blue-eMSTO and red-eMSTO main-sequence populations.}
    \label{eMSTO_spatial}
  \end{figure}

To distinguish between the tidal-locking in binaries and star-disk interactions mechanism, we analyzed the spatial distribution of the eMSTO stars as illustrated in Figure~\ref{eMSTO_spatial}. The tidal locking binary scenario predicts that the slow-rotating blue-eMSTO stars should exhibit preferential spatial concentration in inner region of NGC 2355 due to dynamical mass segregation \citep{2023JApA...44...71M}, as these binary systems would have higher masses than the fast-rotating single red-eMSTO stars \citep{2024MNRAS.532.1212M,2025ApJ...989..123M}. Contrary to the expectations of the tidal-locking theory, we found that the red-eMSTO stars i.e. the fast-rotating stars were predominantly concentrated in the inner region of the cluster as clearly visible in Figure~\ref{eMSTO_spatial}. The cumulative radial distribution plot of the eMSTO stars shown in Figure~\ref{eMSTO_spatial} also exhibit higher fraction of the red-eMSTO stars in the central region of the cluster compared to the slow-rotating blue-eMSTO stars. These observational findings agree with the prediction of \cite{2020MNRAS.495.1978B} for star-disk interaction mechanism that the fast-rotating stars would preferentially located in central region due to higher circumstellar disk distruction in pre-main-sequence phase caused by photoevaporation and dynamical interactions \citep{2021MNRAS.508.3710R,2024AJ....167..120V}. 
\section*{Conclusions} 
Our comprehensive analysis of the eMSTO phenomenon in open cluster NGC 2355 provides evidence that the observed eMSTO is primarily driven by spread rotation rates of the eMSTO stars. The spatial distribution of fast-rotating stars, which are centrally concentrated, rules out the binary tidal-locking scenario. Instead, our findings provide the observational evidence supporting star-disk interactions during the pre-main-sequence phase as the most likely reason for the rotational spread in the eMSTO stars. The early destruction of circumstellar disks in the dense cluster core explains the preferential spin-up of stars in the central region, ultimately manifesting as the eMSTO phenomenon.
\bibliographystyle{iaulike}
\bibliography{manuscript}

\end{document}